\documentclass[11pt,onecolumn,twoside,journal]{article}
\usepackage{amsmath, amssymb, enumerate,graphicx,epsfig}
\usepackage{amsmath}
\usepackage{amsfonts}
\usepackage{enumerate}
\usepackage{epsfig}

\begin{document}

\title{\bf  Group-Server Queues}


\author{\Large Quan-Lin Li$^{a}$, Jing-Yu Ma$^{a}$, Mingzhou Xie$^{b}$ and Li Xia$^{b}$ \\
{\em $^{a}$School of Economics and Management Sciences}\\
{\em Yanshan University, Qinhuangdao 066004 China}\\
liquanlin@tsinghua.edu.cn\\
{\em $^{b}$Center For Intelligent and Networked Systems}\\
{\em Department of Automation and TNList} \\
{\em Tsinghua University, Beijing 100084 China}\\
xial@tsinghua.edu.cn}

\date{}
\maketitle



\begin{abstract}
By analyzing energy-efficient management of data centers, this
paper proposes and develops a class of interesting {\it Group-Server
Queues}, and establishes two representative group-server queues
through loss networks and impatient customers, respectively.
Furthermore, such two group-server queues are given model
descriptions and necessary interpretation. Also, simple mathematical
discussion is provided, and simulations are made to study the
expected queue lengths, the expected sojourn times and the expected
virtual service times. In addition, this paper also shows that this
class of group-server queues are often encountered in many other
practical areas including communication networks, manufacturing
systems, transportation networks, financial networks and healthcare
systems. Note that the group-server queues are always used to design
effectively dynamic control mechanisms through regrouping and
recombining such many servers in a large-scale service system by
means of, for example, bilateral threshold control, and customers
transfer to the buffer or server groups. This leads to the
large-scale service system that is divided into several adaptive and
self-organizing subsystems through scheduling of batch customers and
regrouping of service resources, which make the middle layer of this
service system more effectively managed and strengthened
under a dynamic, real-time and even reward optimal framework. Based on
this, performance of such a large-scale service system may be
improved greatly in terms of introducing and analyzing such
group-server queues. Therefore, not only analysis of group-server
queues is regarded as a new interesting research direction, but there also
exists many theoretical challenges, basic difficulties and open
problems in the area of queueing networks.
\end{abstract}

\section{Introduction}

In this paper, we propose and develop a class of interesting
\textit{Group-Server Queues }by analyzing energy-efficient management of
data centers, and establish two representative group-server queues by means
of loss networks and impatient customers, respectively. Also, we show that this class of
group-server queues are often encountered in many other practical areas including
communication networks, manufacturing systems, transportation networks,
financial networks and healthcare systems. Therefore,
not only analysis of group-server queues is regarded as a
new interesting research direction, but there also
exists many theoretical challenges, basic difficulties and open problems in the area of queueing networks.

Data centers are main infrastructure platforms for various kinds of
large-scale practical information systems, and always offer economies of scale for
network, power, cooling, administration, security and surge capacity. So far
it has been an interesting research direction to reduce the
server farm energy requirements and to optimize the power efficiency which may be
viewed as a ratio of performance improvement to power consumption reduction. For
energy-efficient management of data centers, some authors have dealt with
several key interesting issues, for example, data center network architecture
by Al-Fares \textit{et al}. \cite{Alf:2008}, Guo \textit{et al}. \cite{Guo:2009} and Pries \textit{et al}. \cite{Pri:2011}; green networks and cloud by Kliazovich \textit{et al}.
\cite{Kli:2010}, Mazzucco \textit{et al}. \cite{Maz:2010b}, Gruber and Keller
\cite{Gru:2010}, Goiri \textit{et al}. \cite{Goi:2011}\cite{Goi:2012} with solar
energy, Li \textit{et al}. \cite{Lic:2011} with wind energy, Wu \textit{et
al}. \cite{WuJ:2013} and Zhang \textit{et al}. \cite{Zha:2016}; networks of
data centers by Greenberg \textit{et al}. \cite{Gre:2009a}\cite{Gre:2009b},
Gunaratne \textit{et al}. \cite{Gun:2008}, Shang \textit{et al}.
\cite{Sha:2010}, Kliazovich \textit{et al}. \cite{Kli:2013} and Wang
\textit{et al}. \cite{Wan:2015}; resiliency of data centers by Heller
\textit{et al}. \cite{Hel:2010} and Baldoni \textit{et al}. \cite{Bal:2014};
revenues, cost and performance by Elnozahy
\textit{et al}. \cite{Eln:2002}, Chen \textit{et al}. \cite{Che:2005}, Benson
\textit{et al}. \cite{Ben:2010}, Dyachuk and Mazzucco \cite{Dya:2010},
Mazzucco \textit{et al}. \cite{Maz:2010a} and Aroca \textit{et al}.
\cite{Aro:2014}; analyzing key factors by Greenawalt \cite{Gre:1994} for
hard disks, Chase \textit{et al}. \cite{Cha:2001} for hosting centers (i.e., the previous one of data
center), Guo \textit{et al}. \cite{Guo:2016a} for base station sleeping
control, Guo \textit{et al}. \cite{Guo:2016b} for edge cloud systems, Horvath
and Skadron \cite{Hor:2007} for multi-tier web server clusters, Lim \textit{et
al}. \cite{Lim:2011} for multi-tier data centers, Rivoire \textit{et al}.
\cite{Riv:2008} for a full-system power model, Sharma \textit{et al}.
\cite{Sha:2003} for QoS, Wierman \textit{et al}. \cite{Wie:2009} for processor sharing, and Xu and Li
\cite{XuH:2014} for part execution.

In analyzing energy-efficient management of data centers, queueing
theory and Markov processes are two effective mathematical tools
both from performance evaluation and from optimal control. Up till
now, few papers have applied queueing theory, together with Markov
processes, to performance analysis of data centers with
energy-efficient management. Chen \textit{et al}. \cite{Che:2005}
proposed a queueing model to control energy consumption of service
provisioning systems subject to Service Level Agreements (SLAs).
Nedevschi \textit{et al}. \cite{Ned:2008} demonstrated that
introducing a sleep state to power management in a data center can
save much of the present energy expenditure, and even simple schemes
for sleeping also offer substantial energy savings. Shang \textit{et
al}. \cite{Sha:2010} used network devices to routing service design,
and made the idle network devices to shut down or to put into a
sleep state. Gandhi \textit{et al}. \cite{Gan:2010a} modeled a
server farms with setup cost by means of an $M/M/k$ queueing system,
where each server can be in one of the following states: work, on,
idle, off, or sleep, employed the popular metric of Energy-Response
time Product (ERP) to capture the energy-performance tradeoff, and
gave the first theoretical result on the optimality of server farm
management policies. From a similar analysis to \cite{Gan:2010a},
Gandhi \textit{et al}. \cite{Gan:2010b} used an $M/M/k$ queueing
system but each server has only three states: on, setup and off, and
obtained the distributions of the response time and of the power
consumption. Further, Gandhi and Harchol-Balter \cite{Gan:2011}, Gandhi \textit{et al}.
\cite{Gan:2012} analyzed effectiveness of dynamic power
management in data centers in terms of an $M/M/k$ model, and found
that the dynamic power management policies are very effective when
the setup time is small, the job size is large or the size of the
data center is increasing; in contrast, the dynamic power management
policies are ineffective for small data centers. Schwartz \textit{et
al}. \cite{Sch:2012} and Gunaratne \textit{et al}. \cite{Gun:2008}
provided the energy-efficient mechanism with dual thresholds.
Schwartz \textit{et al}. \cite{Sch:2012} presented a queueing model
to evaluate the trade-off between the waiting time and the energy
consumption, and also developed a queueing model with thresholds to
turn-on reserve servers when needed. Gunaratne \textit{et al}.
\cite{Gun:2008} developed a single-server queue with state-dependent
service rates and with dual thresholds for service rate transitions.

{\bf Contributions of this paper:} The main contributions of this
paper are threefold. The first one is to propose and develop a class
of interesting {\it Group-Server Queues}, and establishes two
representative group-server queues by means of loss networks and
impatient customers, respectively, under a practical background for
analyzing energy-efficient management of data centers. The second
contribution is to set up a general framework for group-server
queues, and to give a detailed discussion for basic issues, for
instance, optimally structural division of server groups, transfer
policies among server groups and/or buffers, dynamic control mechanism design,
revenue management and cost control. The third contribution is to
provide a simple mathematical analysis for a two-group-server loss
queue, and also to design simulation experiments to evaluate the expected
queue lengths, the expected sojourn times and the expected virtual
service times for a three-group-server loss queue and a
three-group-server queue with an infinite buffer. Note that this
class of group-server queues are often encountered in many other
practical areas including communication networks, manufacturing
systems, transportation networks, financial networks and healthcare
systems. Therefore, the methodology and results given in this paper
provide highlights on a new class of queueing networks called
group-server queues, and are applicable to the study of large-scale
service networks in practice.

{\bf Organization of this paper:} The structure of this paper is
organized as follows. In Section 2, we propose and develop a class
of new interesting queueing models: \textit{Group-Server Queues},
and establish a basic framework of the group-server queues, such as,
model structure, operations mechanism, necessary notation and key
factors. In Section 3, we describe a group-server loss queue from
analyzing energy-efficient management of data centers, where the
loss mechanism makes a finite state space so that the group-server
loss queue must be stable. In Section 4, we describe a group-server
queue with impatient customers and with an infinite buffer, where
some key factors of this system are discussed in detail. In Section
5, we provide a simple mathematical analysis for the
two-group-server loss queue, and several open problems are listed.
In Section 6, we design simulation experiments for performance
evaluation of the three-group-server loss queues and of the
three-group-server queues with an infinite buffer, and specifically,
we analyze their expected queue lengths, the expected sojourn times
and the expected virtual service times. Finally, some concluding
remarks are given in Section 7.

\section{A Basic Framework of Group-Server Queues}

In this section, we propose a class of new interesting queueing models:
\textit{Group-Server Queues}, and establish a basic framework for the
group-server queues, for instance, model structure, operations mechanism, necessary
notation and key factors.

We propose such group-server queues by analyzing energy-efficient
management of data centers. As analyzed in the next two sections,
two representative group-server queues are established according to
the need of energy saving. To realize energy saving in data center
networks, a sleep (or off) state introduced to some servers is a
class of key techniques. Using States on, sleep, off and others, a
large-scale service system with more servers is divided into several
subsystems, each of which contains some servers having certain common
attributes. For example, a data center has a set of all different
attributes: $E=\left\{ \text{work, on, sleep, off}\right\}  $, where
`work' denotes that a server is `on' and is also serving a customer;
`on' denotes that a server is idle and ready to serve; `sleep'
denotes that a server is at the dormancy stage with lower power
consumption; and `off' denotes that a server is shut down, where a
setup time may be needed if the server change its state from off to on (or sleep).

Now, we provide a concrete example how to establish the different groups of servers. Let
the set of all the servers in the data center be $\Omega =\left\{ \text{Server 1, Server 2,}\right. $ $\left. \text{Server 3,
\ldots , Server }N\right\} $.
Then the service system of the data center can be divided into
three subsystems whose server groups are given by%
\[
\Omega_{1}=\left\{  \text{Server 1, Server 2, \ldots, Server }n\right\}
,\text{ each server is either at work or on;}%
\]%
\begin{align*}
\Omega_{2} = &  \left\{  \text{Server }n+\text{1, Server }n+\text{2,
\ldots,
Server }n+m\right\}  ,\\
&  \text{each server is either at work, on, or sleep;}%
\end{align*}%
\begin{align*}
\Omega_{3} = &  \left\{  \text{Server }n+m+\text{1, Server
}n+m+\text{2,
\ldots, Server }N\right\}  ,\\
&  \text{each server is either at work, on, sleep or off.}%
\end{align*}
It is seen that the attributes of servers in $\Omega_{2}$ are more than that in $\Omega_{1}$, that is, sleep. While
the sleep attribute makes $\Omega_{2}$ practically different from $\Omega_{1}$. See Fig.1 in the next section for an intuitive understanding. At the same time, to further understand the attribute role played by the energy efficient mechanism of data centers, readers may refer to the next two sections for establishing and analyzing two representative group-server queues.

In general, this class of group-server queues are often encountered in many other
practical areas, for example, communication networks for green and energy
saving, manufacturing systems for priority use of high-price devices,
transportation networks with different crowded areas, and healthcare systems
having different grade hospitals. Therefore, analyzing the group-server queues is an
interesting research direction both in the queueing area and in many practical
fields such as computer, communication, manufacturing, service, market,
finance and so on. To our best knowledge, no pervious work has looked at and summarized the
group-server queues from a theoretical or practical framework yet.

Based on the above analysis, we provide a basic framework for the group-server
queues, and discuss model structure, operations mechanism, necessary
notation and key factors as follows:

\textbf{Server Groups:} We assume that a large-scale service system contains
many servers whose set is given by
\[
\Omega=\left\{  \text{Server 1, Server 2, Server 3, \ldots, Server
}N\right\},
\]
and they also have some different attributes whose set is given by%
\[
E=\left\{  \text{A}_{1}\text{, A}_{2}\text{, A}_{3}\text{, \ldots, A}%
_{r}\right\}  ,
\]
where $A_i$ is an attribute for $1\leq i\leq r$. From practical need and physical behavior of a large-scale service system, the
attribute set $E$ is divided into some different subsets as follows:
\[
E=E_{1}\cup E_{2}\cup\cdots\cup E_{s},\text{ \ }s\leq r.
\]
Note that $E_{i}$ and $E_{j}$ may have common elements for $1\leq i< j\leq s$. Applying the subsets $E_{i}$ for $1\leq i\leq s$ to system behavior, the server set $\Omega$ is
divided into some different groups or subsets as follows:%
\[
\Omega=\Omega_{1}\cup\Omega_{2}\cup\cdots\cup\Omega_{s},
\]
where the server group $\Omega_{i}$ well corresponds to the attribute subset $E_{i}$ for $1\leq i\leq r$. Note that the server groups $\Omega_{i}$ for $1\leq i\leq s$ are disjoint. In this case, the large-scale service system is divided into $s$
subsystems, each of which contains the servers in the group
$\Omega_{i}$ having certain common attributes in one of the subsets $E_{i}$ for
$1\leq i\leq s$.

In practice, the attributes A$_{i}$ for $1\leq i\leq r$ have a wide range of
meanings, for instance, states, properties, behaviors, control and mechanism. From such a setting, it is clear that the server grouping of a large-scale service system should not be unique.

\textbf{Arrival processes:} In the group-server queue, we assume
that customer arrivals at this system are a renewal process with
stationary arrival rate $\lambda$. For the customer arrivals, we shall
encounter two issues: (a) Routing allocation mechanism, for example,
joining the shortest queue length, and joining the shortest sojourn time. (b) Arrival
rate control, for example, arrival rates depending on system states,
arrival rates depending on prices, arrival rates depending on
sojourn times, and arrival rates depending on threshold control.

\textbf{Service processes:} In the $i$th server group, the service times of
customers are i.i.d. with general distribution function $F_{i}\left(
x\right)  $ of stationary service rate $\mu_{i}$ for $1\leq i\leq N$.
For the customer service processes, we shall encounter two issues: (a) service
disciplines, for example, FCFS, LCFS, processor sharing, and matching service.
(b) Service rate control, for example, service rates depending on system
states, service rates depending on prices, service rates depending on sojourn
times, and service rates depending on threshold control.

\textbf{Customers transfer among server groups: }The customers in heavy-load
server groups are encouraged to transfer into light-load server groups; the
customers in low-speed-service server groups are encouraged to transfer into
high-speed-service server groups; the customers in high-cost-service server
groups are encouraged to transfer into low-cost-service server groups; and so
forth. Under customer transferring, the residual service times of non-exponential distributions always make model analysis more difficult
and challenging.

\vskip 0.5cm

\textbf{Stability is a difficult issue: }Since the group-server queue is always
a large-scale complicated stochastic system, its stability and associated conditions are always
very difficult to study. To easily deal with system stability, it is a simple method to
introduce loss networks or impatient customers to the group-server queues,
where the former is to use the finite state space, while the latter is to
apply stability of the renewal processes. Therefore, this paper uses the loss
networks and the impatient customers to set up some examples, which show how to simply guarantee stability of some group-server queues.

\section{Group-Server Loss Queues}

In this section, we describe a group-server loss queue from
analyzing energy-efficient management of data centers. Note that the
`Loss Mechanism' is to set up a finite state space, whose purpose is to
guarantee stability of the group-server loss queue.

In the energy-efficient management study of data center networks, it was an
effective way to introduce two states: sleep and off for some
servers. Based on this, we can make some different states: work, on,
sleep, off and others, and the servers of the data center are
grouped as
$\Omega=\Omega_{0}\cup\Omega_{1}\cup\cdots\cup\Omega_{N}$.
Concretely, a simple group-server loss queue is constructed under
two states: work-on, and sleep. Here, work and on are assumed to
have the same power consumption, hence work and on are regarded as a
state: work-on. See Fig.1, for understanding the $N+1$ server groups and even for intuitively understanding the group-server loss queue.

\begin{figure}[htbp]
\vspace{6mm}
\begin{center}
\includegraphics[width=15 cm]{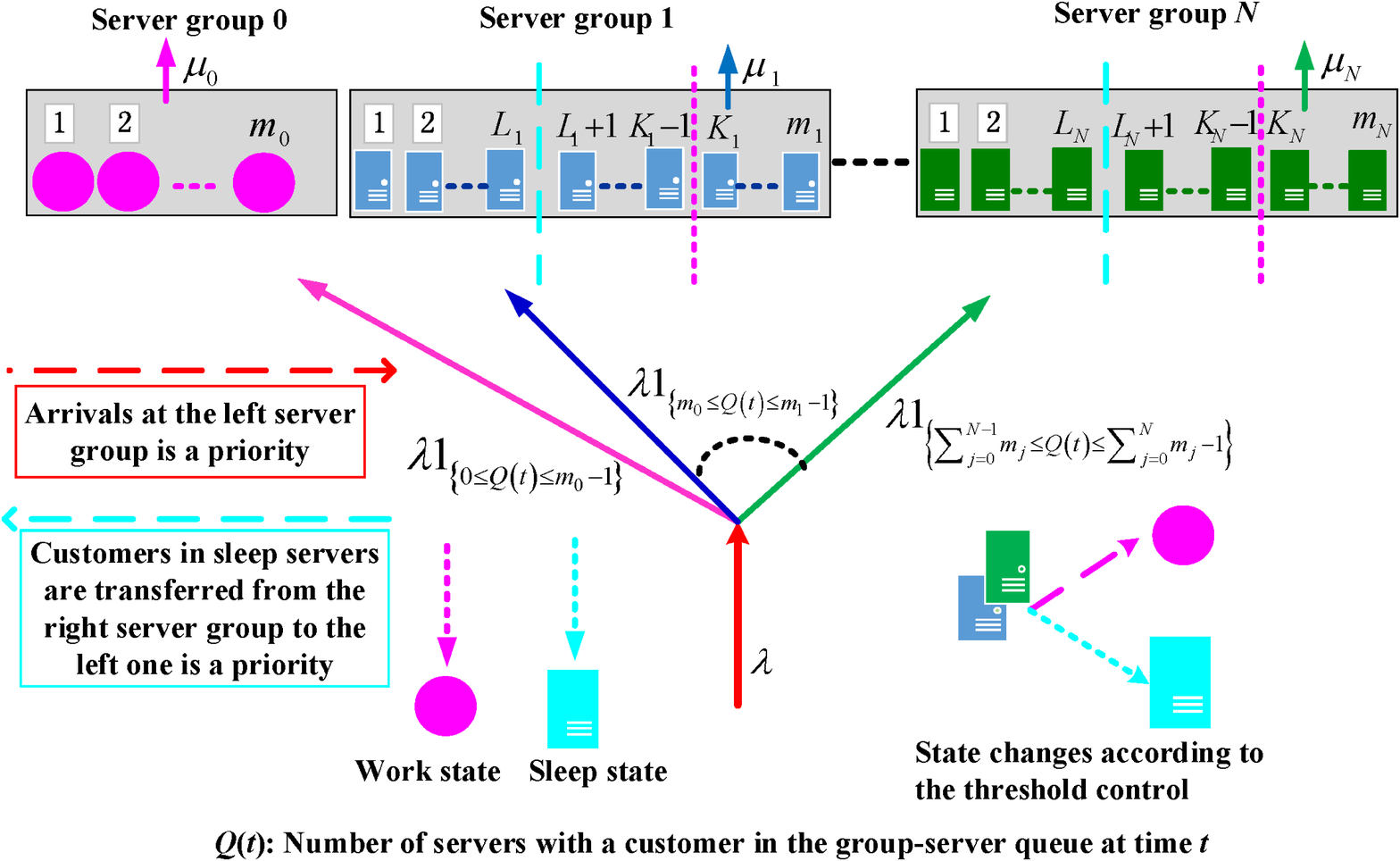}
\caption{Physical structure of the group-server loss queue }\label{energysavingp}
\end{center}
\end{figure}

Now, we use the data center to describe a group-server loss queue, and set up energy
efficient mechanism, system parameters and model notation as follows:

\textbf{(1) Server groups:} We assume that the data center contains
$N+1$ different server groups, each of which is a subsystem of the
data center. For $j\in\left\{  0,1,2,\ldots,N\right\}  $, the
$j$th server group contains $m_{j}$ same servers. Thus the data
center contains $\sum_{j=0}^{N}m_{j}$ servers in total. Note that
the $N+1$ different server groups can be divided into two basic
categories: (a) Server group $0$ is special, because its each server
has only one state: work-on. Hence server group $0$ with $m_{0}$
same servers is viewed as the base-line group in the data center.
(b) Each server of the other $N$ server groups has two states:
work-on and sleep.

\textbf{(2)} \textbf{Arrival processes:} The arrivals of customers
at the data center from outside are a renewal process with
stationary arrival rate $\lambda$. An arriving customer
preferentially enters one idle server of the leftmost server group
with idle servers. We assume that each server and the data center all have
no waiting room, while each server receives and serves only one customer at a time. Hence any new arrival is lost once all the
servers contain their one customer, that is, the system is full when it has
at most $\sum _{j=0}^{N}m_{j}$ customers.

\textbf{(3)} \textbf{Service processes: }In the $j$th server group, the
service times of customers are i.i.d. with general distribution function
$F_{j}\left(  x\right)  $ of stationary service rate $\mu_{j}$ for
$0\leq j\leq N$.

\textbf{(4) Bilateral threshold control: }Except server group $0$,
each server in the other $N$ server groups have two states: work-on
and sleep. To switch between work-on and sleep, it is necessary for
the $j$th server group to introduce a positive integer pair $\left(
L_{j},K_{j}\right)  $ with $0\leq L_{j}\leq K_{j}\leq m_{j}$, which
leads to a class of bilateral threshold control applied to
energy-efficient management of the data center networks..

To realize energy saving effectively, using the two states: work-on
and sleep, a \textit{bilateral threshold control} is introduced as
follow: Once there are more than $K_{j}$ customers in the $j$th
server group, then each server of the $j$th server group immediately
enters State work-on. On the contrary, when there are less than
$L_{j}$ customers in the $j$th server group, then each server of the
$j$th server group immediately enters State sleep. Thus for the data
center, we have the coupled threshold control parameters as follows:
\[
\left\{  \left(  L_{j},K_{j}\right)  :0\leq L_{j}\leq K_{j}\leq m_{j},1\leq
j\leq N\right\}  .
\]

\textbf{(5) Customers concentratively transfer among the server groups:} In
order to make the sleep servers to enter State work-on as soon as
possible, it is necessary to concentratively transfer those customers in the sleep
servers of the rightmost server group with sleep servers into the
idle servers (on or sleep) of the leftmost server group. Using such a
way, this maximizes the number of customers in the leftmost server
group with sleep servers such that the number of servers with a customer fast goes to over the
integer $K_{j}$, which leads to that the sleep servers is started up
and enters State work-on. In this case, the most sleep servers in
the data center are setup at State work-on so that more and more
customers are served as soon as possible.

\textbf{(6) Energy consumption:} We assume that the power consumption rates
for the $1+N$ server groups are given by $P_{W_{0}},P_{W_{1}},\ldots,P_{W_{N}%
}$ for State work-on; and
$P_{S_{0}}=0,P_{S_{1}},P_{S_{2}},\ldots,P_{S_{N}}$ for State sleep.
To realize energy saving, let $0<P_{S_{j}}<P_{W_{j}}$ for
$j=1,2,\ldots,N$.

We assume that all the random variables in the system defined above are
independent of each other.

\vskip 0.5cm

{\bf A basic issue:} Establishing some cost (or reward) functions is to
evaluate a suitable trade-off between the sojourn time and the energy consumption.
To concentratively reduce the sojourn time and to save energy, some effective methods are proposed and developed,
such as, (a) the bilateral threshold control, (b) the customers at sleep servers concentratively transfer among the server groups,
and (c) the residual service times are wasted or re-used. We need to analyze their performance and trade-off
due to some mutual contradiction between reducing the sojourn time and saving the energy.

\section{Group-Server Queues with Impatient Customers}

In this section, we consider a group-server queue with impatient customers,
which is refined and abstracted from energy-efficient management of data center
networks, where the impatient customers are introduced to guarantee stability of this
group-server queue with an infinite buffer.

In the data center, we still introduce two states: work-on and sleep for some
servers. Based on the two different states, the servers of the data center are
grouped as $\Omega=\Omega_{0}\cup\Omega_{1}\cup\Omega_{2}\cup\cdots\cup
\Omega_{N}$. See Fig.2 both for the $N+1$ server groups and for the
group-server queue with impatient customers and with an infinite buffer.

Now, we describe a group-server queue with impatient
customers and with an infinite buffer, and also explain energy
efficient mechanism, system parameters and model notation as
follows:
\begin{figure}[htbp]
\begin{center}
\includegraphics[width=15 cm]{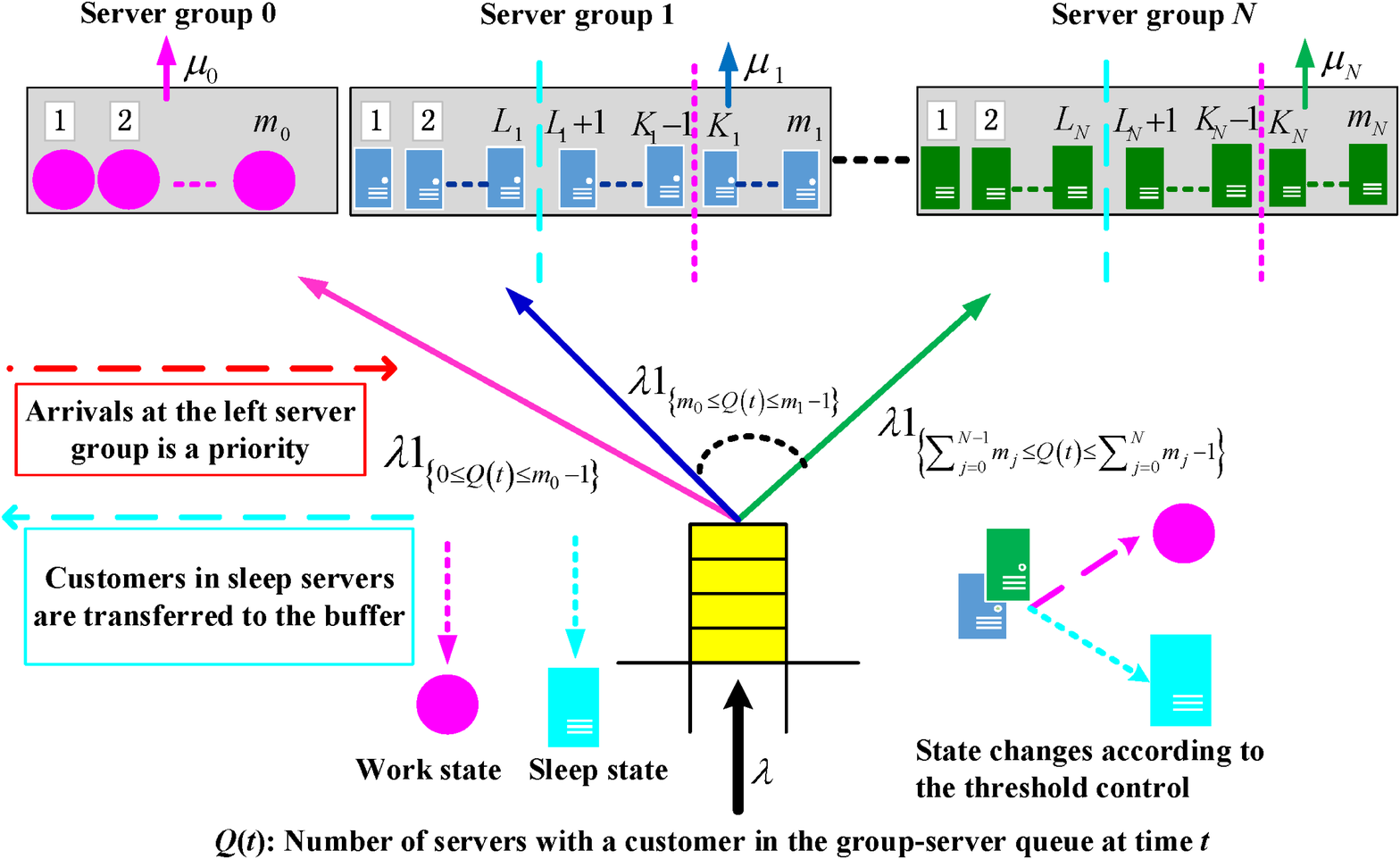}
\caption{Physical structure of the group-server queue with  impatient customers}\label{energysavingp}
\end{center}
\end{figure}

\textbf{(1) Server groups:} We assume that the data center contains
$N+1$ different server groups, each of which is a subsystem of the
data center. For a $j\in\left\{  0,1,2,\ldots,N\right\}  $, the
$j$th server group contains $m_{j}$ same servers. Thus the data
center contains at most $\sum_{j=0}^{N}m_{j}$ servers. Note that the
$N+1$ different server groups can be divided into two basic
categories: (a) Server group $0$ is special, because its each server
has only one state: work-on, hence server group $0$ with $m_{0}$
same servers is viewed as the base-line group in the data center.
(b) Each server of the other $N$ server groups has two states:
work-on and sleep.

\textbf{(2)} \textbf{Arrival processes:} The arrivals of customers at the data
center from outside are a renewal process with stationary arrival rate
$\lambda$. Each arrival customer must first enter the infinite buffer, then he
is assigned to the $N+1$ server groups according to the following allocation rules:

(2-A1) Each server in server group $0$ is always at State work-on. If
server group $0$ have some idle servers, then the arriving customer
can immediately enter one idle server in server group $0$ and then
receive his service.

(2-A2) Each server in server group $1$ is at State sleep. If server group
$0$ does not exist any idle server, then the new arrival customers have to
queue and wait in the buffer. Once the number of customers waiting in the
buffer is not less than $K_{1}$, then each server in server group $1$ is
started up to State work-on from State sleep, and all the customers in the buffer but at
most $m_{1}$ customers enter server group $1$ and then receive their service.

(2-A3) Each server in server group $2$ is at State sleep. If server
groups $0$ and $1$ do not exist any idle server, then the new arrival
customers have to queue and wait in the buffer. Once the number of customers
waiting in the buffer is not less than $K_{2}$, then each server in server
group $2$ is started up to State work-on from State sleep, and all the
customers in the buffer but at most $m_{2}$ customers enter server group $2$
and then receive their service.

(2-A4) Let $2\leq l\leq N-1$. Each server in server group $l+1$ is at
State sleep. If server groups $0,1,2,\ldots,l$ do not have any idle server,
then the new arriving customers have to queue and wait in the buffer. Once the
number of customers waiting in the buffer is not less than $K_{l+1}$, then
each server in server group $l+1$ is started up to State work-on from State
sleep, all the customers in the buffer but at most $m_{l+1}$ customers enter
server group $l+1$ and then receive their service.

(2-B) If the $N+1$ server groups do not exist any idle server, then the new
arriving customers have to queue and wait in the buffer.

(2-C) In the $N+1$ server groups, if there exists one idle server whose state
is work-on, then an arriving customer in the buffer will immediately enter this
server and then receive his service.

\textbf{(3)} \textbf{Service processes: }In the $j$th server group, the
service times of customers are i.i.d. with general distribution function
$F_{j}\left(  x\right)  $ of stationary service rate $\mu_{j}$ for
$0\leq j\leq N$.

\textbf{(4) Customer impatient processes: }Each customer in this system has
an exponential patient time of impatient rate $\theta$.

\textbf{(5) Bilateral threshold control: }Except server group $0$, each server
in the other $N$ server groups have two states: work-on and sleep. To switch
between work-on and sleep, it is necessary for the $j$th server group to
introduce a positive integer pair $\left(  L_{j},K_{j}\right)  $ with $0\leq
L_{j}\leq K_{j}\leq m_{j}$, which leads to a bilateral threshold control by
means of energy-efficient management of the data center networks.
By using the two states: work-on and sleep, a \textit{Bilateral Threshold
Control} is introduced as follow:

(5-1) \textbf{From sleep to work-on: }We assume that each server in server
group $j$ is at State sleep, and an idle server does not exist in server groups $0,1,2,\ldots,j-1$
, for $1\leq j\leq N$. If there are not less than $K_{j}$
customers in the buffer, then each server in server group $j$ is started up to
State work-on from State sleep, all the customers in the buffer but at most $m_{j}$
customers enter server group $j$ and then receive their service.

(5-2) \textbf{From work-on to sleep: }We assume that each server in server
group $j$ is at State work-on. If there are less than $L_{j}$ customers being
served in server group $j$, then each
server of server group $j$ immediately enters State sleep from State work-on,
and those customers being served at server group $j$ are transferred to the
head of the buffer. For such transferred customers, we assume that those
service times obtained by the transferred customers, will become zero, and their
service get start again (note that another useful case is that the received service times can be cumulative, once arriving at the total service time, the service is completed) immediately. Also, to allocate the customers in the buffer to some server
groups, each customer transferred in the head of the buffer is the same as
the new arriving customers. Thus we have the coupled threshold control parameters%
\[
\left\{  \left(  L_{j},K_{j}\right)  :0\leq L_{j}\leq K_{j}\leq m_{j},1\leq
j\leq N\right\}  .
\]

\textbf{(6) Customers transfer to the buffer:} In order to save energy
as much as possible, if there are less than $L_{j}$ customers being served in
server group $j$, then it is necessary to transfer those customers in the
server group $j$ at State work-on into the head of the buffer, and each server
of server group $j$ immediately enters State sleep from State work-on.

\textbf{(7) Energy consumption:} We assume that the power consumption rates
for the $1+N$ server groups are given by $P_{W_{0}},P_{W_{1}},\ldots,P_{W_{N}%
}$ for State work-on; and
$P_{S_{0}}=0,P_{S_{1}},P_{S_{2}},\ldots,P_{S_{N}}$ for State sleep.
For energy saving, let $0<P_{S_{j}}<P_{W_{j}}$ for $j=1,2,\ldots,N$.

We assume that all the random variables in the system defined above are
independent of each other.

\vskip 0.5cm

{\bf A basic issue:} Constructing some cost (or reward) functions is
to evaluate a suitable trade-off between the sojourn time and the
energy consumption. Note that some effective methods are proposed
and developed, such as, (a) the bilateral threshold control, (b) customers in sleep servers transfer to the buffer, and (c) residual
service times are wasted or re-used. We need to analyze their performance and the suitable
trade-off due to some mutual contradiction between reducing the
sojourn time and saving the energy.

\vskip 0.5cm

{\bf Further discussion for stability:} For the group-server queue
with an infinite buffer drawn in Fig.2, it is easy to give a
sufficient condition of system stability: $\rho=\lambda / m_0
\mu_{0}<1$, by means of a path coupling or comparison of a Markov
process.

It is seen from Fig.2 that server groups $1, 2, \ldots, N$
provide many service resources or ability in processing the
customers, and this buffer also plays a key role in concentratively transferring the
customers in sleep servers to improve service ability of the whole system.
Therefore, it may be an interesting open problem to set up the necessary
conditions under which the system stability is, how to be influenced
by the key factors or parameters, as follows: (a) the bilateral threshold control, (b) customers in sleep servers transfer to the buffer, and (c) many residual service times are wasted or re-used.

\section{Mathematical Analysis and Open Problems}

In this section, we provide some simple mathematical analysis for a
two-group-server loss queue with server group $0$ and server group $1$, a
whole detailed investigation of which was given in Li \textit{et al.}
\cite{Li:2017}. From such mathematical analysis, it is seen that analyzing
more general group-server queues is interesting, challenging and difficult,
and thus several open problems are listed for the future study of
group-server queues.

\subsection{Some mathematical analysis}

\textbf{A special model:} $N=1$. We consider a special
two-group-server loss queue with server group $0$ and server group
$1$. See model descriptions for a more general case given in Section
3 in more details.

For convenience of readers, it is still necessary to simply re-list
some assumptions and descriptions for this special case as follows:
\textbf{(a)} Server group $0$ contains $n$ servers, each of which
has only one state: work-on; while server group $1$ contains $m$
servers, each of which has two different states: work-on and sleep.
\textbf{(b)} No waiting room is available both at each server and in the group-server queue. Once there are $n+m$
customers in this systems (i.e., one server is serving a customer), any new
customer arrival will be lost due to no waiting room. \textbf{(c)}
The arrivals of customers at this two-group-server queue are a
Poisson process with arrival rate $\lambda$. \textbf{(d)} The
service times provided by server group $0$ and by server group $1$
are i.i.d. with two exponential distributions of service rate
$\mu_{0}$ and $\mu_{1}$, respectively. \textbf{(e)}
\textit{Unilateral threshold control: }$L_{1}=0$ and $K_{1}>0$. In
this case, if server group $0$ are full with $n$ serving customers,
and if there are not less that $K_{1}$ customers waiting at server group $1$ with $m$ sleep servers, then each server
of server group $1$ immediately enters State work-on, and then
provides service for its customer. \textbf{(f)} Because of
$L_{1}=0$, once all the customers in server group $1$ have completed
their service, the $m$ servers of server group $1$ immediately
switch to State sleep from State work-on; if there are some idle
(on) servers in server group $0$, and if there are customers waiting
in server group $1$ in which each server is at State sleep, then the
customers waiting at the sleep servers of server group $1$ can
transfer to those idle (on) servers of server group $0$. \textbf{(g)
}For each server in the two server groups, the power consumption
rates are listed as: $P_{W_{0}}$ and $P_{W_{1}}$ for State work-on,
and $P_{S_{1}}$ for State sleep ($P_{S_{0}}=0$ because each server in server
group $0$ does not have State sleep). To save energy, we assume that
$0<P_{S_{1}}<P_{W_{1}}$.

\textbf{A QBD process:} For this two-group-server loss queue, the states of
the corresponding Markov process are defined as the tuples: $\left(
l_{0},i;l_{1},j\right)  $ as shown in Fig.3, where $l_{0}$ $=W$,
$l_{1}\in\left\{  W,S\right\}  $, $W$ and $S$ denote States work-on and sleep,
respectively. Let $i$ be the number of servers serving a customer in
server group $0$, and $j$ the number of servers serving a customer in
server group $1$. It is seen from Fig.3 that for $\left(  W,i;l_{1}%
,j\right)  $, we can set up four different sets as follows:%

\begin{figure}[ptb]
\centering    \includegraphics[width=21cm, angle=270]{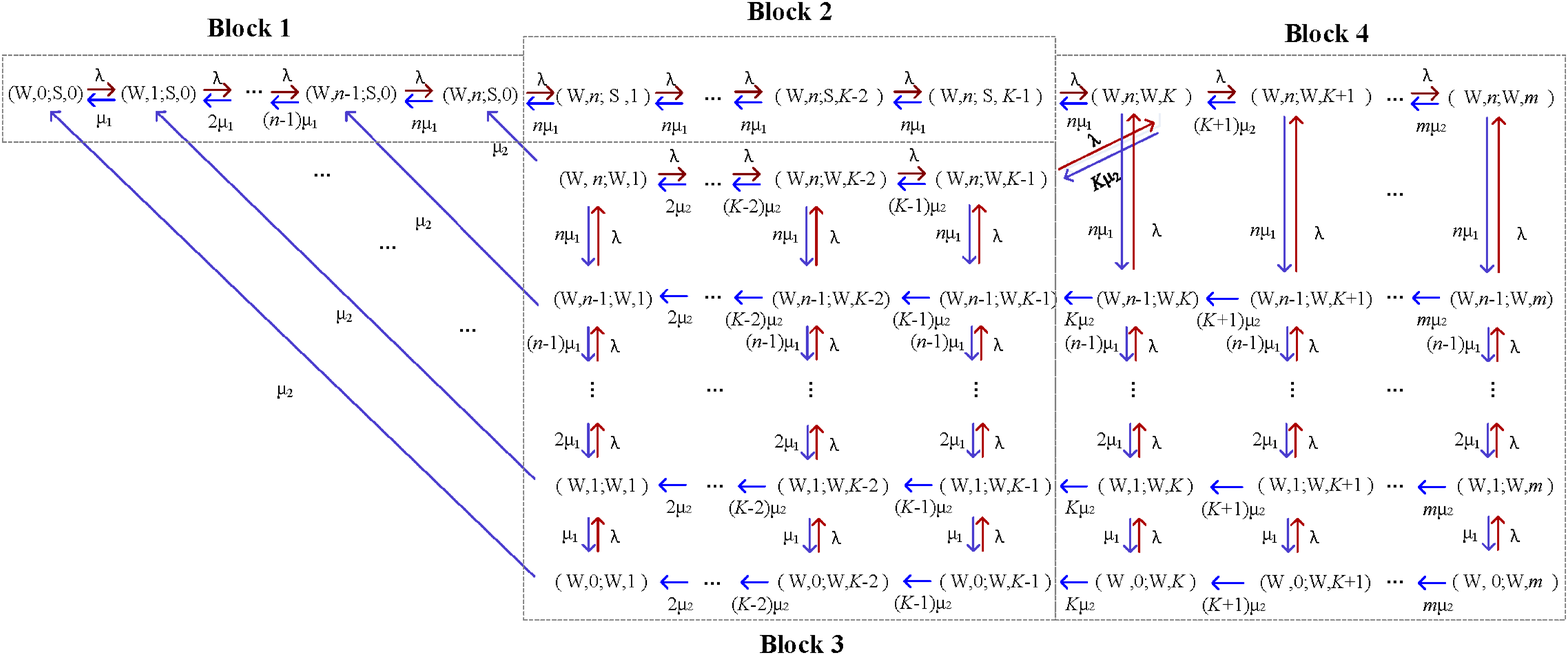}
\newline \caption{State transition relations of Markov process in a
two-group-server loss queue}%
\label{figure:fig2}%
\end{figure}

\[%
\begin{array}
[c]{lllll}%
\text{(1)} & \text{In Block 1,} & i\in\left\{  0,1,\ldots,n-1,n\right\}  ; &
l_{1}=S, & j=0;\\
\text{(2)} & \text{in Block 2,} & i\in\left\{  n\right\}  ; & l_{1}=S, &
j\in\left\{  1,2,\ldots,K-1\right\}  ;\\
\text{(3)} & \text{in Block 3,} & i\in\left\{  0,1,\ldots,n-1,n\right\}  ; &
l_{1}=W, & j\in\left\{  1,2,\ldots,K-1\right\}  ;\\
\text{(4)} & \text{in Block 4,} & i\in\left\{  0,1,\ldots,n-1,n\right\}  ; &
l_{1}=W, & j\in\left\{  K,K+1,\ldots,m\right\}  .
\end{array}
\]

We denote by $N_{0}\left(  t\right)  $ and $N_{1}\left(  t\right)  $
the numbers of servers serving a customer in server group $0$
and server group $1$, respectively; and $S_{0}\left(  t\right)$ and $S_{1}\left(  t\right)$ the
states of servers in server groups 0 and 1, where $S_{0}\left(  t\right)=W$ and $S_{1}\left(  t\right)
\in\left\{ W,S\right\}$. Let
$\mathbf{X}\left( t\right) =\left(S_{0}\left(  t\right) ,N_{0}\left( t\right) ;S_{1}\left(
t\right) ,N_{1}\left( t\right) \right)  $ with $S_{0}\left(  t\right)=W$. Then $\left\{
\mathbf{X}\left( t\right) ,t\geq0\right\} $ is a QBD process with
finitely many levels. From Fig.3, it is seen that the QBD process
$\left\{ \mathbf{X}\left( t\right) ,t\geq0\right\} $ has a state
space
$\mathbf{\Omega}=\mathbf{\Omega}_{0}\cup\mathbf{\Omega}_{1}\cup\mathbf{\Omega
}_{2}\cup\mathbf{\cdots\cup\Omega}_{m}$, where
\begin{align*}
\mathbf{\Omega}_{0} &  =\left\{  \left(  W,0;S,0\right)  ,\left(
W,1;S,0\right)  ,\ldots,\left(  W,n-1;S,0\right)  ,\left(  W,n;S,0\right)
\right\}  ,\\
\mathbf{\Omega}_{1} &  =\left\{  \left(  W,n;S,1\right)  \right\}
\cup\left\{  \left(  W,n;W,1\right)  ,\left(  W,n-1;W,1\right)  ,\ldots
,\left(  W,0;W,1\right)  \right\}  ,\text{ }\\
\mathbf{\Omega}_{2} &  =\left\{  \left(  W,n;S,2\right)  \right\}
\cup\left\{  \left(  W,n;W,2\right)  ,\left(  W,n-1;W,2\right)  ,\ldots
,\left(  W,0;W,2\right)  \right\}  ,\\
&  \text{ \ \ \ \ \ }\vdots\text{
\ \ \ \ \ \ \ \ \ \ \ \ \ \ \ \ \ \ \ \ \ \ \ \ \ }\vdots\\
\mathbf{\Omega }_{K-1}& =\left\{ \left( W,0;S,K-1\right) \right\} \cup
\left\{ \left( W,n;W,K-1\right) ,\left( W,n-1;W,K-1\right) ,\ldots ,\right.
\\
& \text{ \ \ \ }\left. \left( W,0;W,K-1\right) \right\} ; \\
\mathbf{\Omega }_{K}& =\left\{ \left( W,n;W,K\right) ,\left(
W,n-1;W,K\right) ,\ldots ,\left( W,1;W,K\right) ,\left( W,0;W,K\right)
\right\} ; \\
\mathbf{\Omega }_{K+1}& =\left\{ \left( W,n;W,K+1\right) ,\left(
W,n-1;W,K+1\right) ,\ldots ,\left( W,0;W,K+1\right) \right\} ;\\
&  \text{ \ \ \ \ \ }\vdots\text{
\ \ \ \ \ \ \ \ \ \ \ \ \ \ \ \ \ \ \ \ \ \ \ \ \ }\vdots\\
\mathbf{\Omega}_{m} &  =\left\{  \left(  W,n;W,m\right)  ,\left(
W,n-1;W,m\right)  ,\ldots,\left(  W,1;W,m\right)  ,\left(  W,0;W,m\right)
\right\}  .
\end{align*}

Let the subset $\mathbf{\Omega}_{j}$ be Level $j$. Then each element in Level $j$ or the
subset $\mathbf{\Omega}_{j}$ is \textit{a phase}. It is clear that the QBD
process $\left\{  \mathbf{X}\left(  t\right)  ,t\geq0\right\}  $ has the
infinitesimal generator as follows:

\[
Q=\left(
\begin{array}
[c]{ccccc}%
\text{ \ }Q_{0,0}\text{\ \ } & \text{\ }Q_{0,1\text{\ \ }} & \text{
\ \ \ \ \ } &  & \\
Q_{1,0} & Q_{1,1} & Q_{1,2} &  & \\
& Q_{2,1} & Q_{2,2} & Q_{2,3} & \\
& \text{ \ \ \ \ \ }\ddots & \text{ \ \ \ \ \ }\ddots & \text{ \ \ \ \ \ }%
\ddots & \\
&  & Q_{m-1,m-2} & Q_{m-1,m-1} & Q_{m-1,m}\\
&  &  & Q_{m,m-1} & Q_{m,m}%
\end{array}
\right)  .
\]

\textbf{A Markov reward process:} Using the power consumption rates, we define
$f(x)$ as an instantaneous reward of the QBD process $\left\{  \mathbf{X}%
\left(  t\right)  ,\text{ }t\geq0\right\}  $ at the state $\mathbf{X}\left(
t\right)  =x$. It is clear that for $x=\left(  W,i;l_{1},j\right)  $ with
$i\in\{0,1,...,n-1,n\},$
\begin{equation}
f\left(  W,i;l_{1},j\right)  =\left\{
\begin{array}
[c]{ll}%
nP_{W_{0}}+mP_{S_{1}}, & l_{1}=S, j\in\{0,1,...,K-1\},\\
nP_{W_{0}}+mP_{W_{1}}, & l_{1}=W, j\in\{1,2,...,m\}.
\end{array}
\right.  \label{gs-3}%
\end{equation}
For simplification of description, we write%
\[
f_{W,i;l_{1},j}=f\left(  W,i;l_{1},j\right)  .
\]
It is seen from Fig.3 and the state space $\mathbf{\Omega}=\mathbf{\Omega
}_{0}\cup\mathbf{\Omega}_{1}\cup\mathbf{\cdots\cup\Omega}_{m}$ that for Level
$0$,%
\[
f_{S,0}=\left(  f_{W,0;S,0},f_{W,1;S,0},\ldots,f_{W,n;S,0}\right)  ;
\]
for Level $j$ with $1\leq j\leq K-1,$%
\[
f_{SW,j}=\left(  f_{W,n;S,j};f_{W,0;W,j},f_{W,1;W,j},\ldots,f_{W,n;W,j}%
\right)  ;
\]
and for\ Level $j$ with $K\leq j\leq m,$%
\[
f_{W,j}=\left(  f_{W,0;W,j},f_{W,1;W,j},\ldots,f_{W,n;W,j}\right)  .
\]
We write
\[
f_{j}=\left\{
\begin{array}
[c]{ll}%
f_{S,0}, & j=0,\\
f_{SW,j}, & 1\leq j\leq K-1,\\
f_{W,j}, & K\leq j\leq m,
\end{array}
\right.
\]
and%
\begin{equation}
f=\left(  f_{0},f_{1},\ldots,f_{K-1},f_{K},\ldots,f_{m}\right)  ^{T}%
.\label{gs-4}%
\end{equation}

Now, for the two-group-server loss queue, we analyze some interesting
performance measures: (a) The expected instantaneous power consumption rate
$E\left[  f\left(  \mathbf{X}\left(  t\right)  \right)  \right]  $; and (b)
the cumulative power consumption $\Phi\left(  t\right)  $ during the time
interval $\left[  0,t\right)  $. At the same time, the probability distribution and the
first passage time involved in the energy saving are also computed in detail.

\textbf{The expected instantaneous power consumption rate:} We note that
$f(\mathbf{X}\left(  t\right)  )$ is the instantaneous power consumption rate
of the two-group-server queue at time $t\geq0$. If the QBD process $Q$ is
stable, then
\[
\underset{t\rightarrow+\infty}{\lim}E\left[  f\left(  \mathbf{X}\left(
t\right)  \right)  \right]  =\underset{\left(  W,i;l_{1},j\right)
\in\mathbf{\Omega}}{\sum}\pi_{W,i;l_{1},j}f_{W,i;l_{1},j}=\pi f,
\]
where $\pi$ is the stationary probability vector of the QBD process $Q$, and
it can be obtained through solving the system of linear equations $\pi Q=0$
and $\pi e=1$ by means of the $RG$-factorizations given in Li \cite{Li:2010},
where $e$ is a column vector of ones.

\textbf{The cumulative power consumption }$\Phi\left(  t\right)  $\textbf{
during the time interval }$\left[  0,t\right)  $

Based on the instantaneous power consumption rate $f\left(  \mathbf{X}\left(
t\right)  \right)  $, we define the cumulative power consumption during the
time interval $\left[  0,t\right)  $ as
\[
\Phi\left(  t\right)  =\int_{0}^{t}f\left(  \mathbf{X}\left(  u\right)
\right)  \text{d}u.
\]

\textbf{(i) Computing the probability distribution:} Now, we compute
the probability distribution of the cumulative power consumption
$\Phi\left( t\right)  $ by means of a partial differential equation
whose solution can explicitly be given in terms of the Laplace and
Laplace-Stieltjes transforms. Let
\[
\Theta\left(  t,x\right)  =P\left\{  \Phi\left(  t\right)  \leq x\right\}
\]
and%
\[
H_{W,i;l_{1},j}\left(  t,x\right)  =P\left\{  \Phi\left(  t\right)  \leq
x,\text{ }\mathbf{X}\left(  t\right)  =\left(  W,i;l_{1},j\right)  \right\}  .
\]
We write that for Level $0$,%
\[
H_{S,0}\left(  t,x\right)  =\left(  H_{W,0;S,0}\left(  t,x\right)
,H_{W,1;S,0}\left(  t,x\right)  ,\ldots,H_{W,n;S,0}\left(  t,x\right)
\right)  ;
\]
for Level $j$ with $1\leq j\leq K-1$,%
\[
H_{SW,j}\left(  t,x\right)  =\left(  H_{W,n;S,j}\left(  t,x\right)
;H_{W,0;W,j}\left(  t,x\right)  ,H_{W,1;W,j}\left(  t,x\right)  ,\ldots
,H_{W,n;W,j}\left(  t,x\right)  \right)  ;
\]
and for Level $j$ with \ $K\leq j\leq m$,%
\[
H_{W,j}\left(  t,x\right)  =\left(  H_{W,0;W,j}\left(  t,x\right)
,H_{W,1;W,j}\left(  t,x\right)  ,\ldots,H_{W,n;W,j}\left(  t,x\right)
\right)  .
\]
Based on this, we write
\[
H_{j}\left(  t,x\right)  =\left\{
\begin{array}
[c]{ll}%
H_{S,0}\left(  t,x\right)  , & j=0,\\
H_{SW,j}\left(  t,x\right)  , & 1\leq j\leq K-1,\\
H_{W,j}\left(  t,x\right)  , & K\leq j\leq m,
\end{array}
\right.
\]
and
\[
\text{ }H\left(  t,x\right)  =\left(  H_{0}\left(  t,x\right)  ,H_{1}\left(
t,x\right)  ,\ldots,H_{K-1}\left(  t,x\right)  ,H_{K}\left(  t,x\right)
,\ldots,H_{m}\left(  t,x\right)  \right)  .
\]
It is clear that
\[
\Theta\left(  t,x\right)  =H\left(  t,x\right)  e.
\]

For a column vector $a=\left(  a_{1},a_{2},\ldots,a_{r}\right)  $ of size $r$,
we write
\[
\Delta\left(  a\right)  =\text{diag}\left(  a_{1},a_{2},\ldots,a_{r}\right)
.
\]
Obviously for the column vector $f$, we have
\[
\Delta=\text{diag}\left(  \Delta\left(  f_{0}\right)  ,\Delta\left(
f_{1}\right)  ,\ldots,\Delta\left(  f_{K-1}\right)  ,\Delta\left(
f_{K}\right)  ,\ldots,\Delta\left(  f_{m}\right)  \right)  .
\]

For the Markov reward process $\left\{  \Phi\left(  t\right)
,t\geq0\right\} $, it follows from Section 10.2 of Chapter 10 in Li
\cite{Li:2010} that the vector function $H\left(  t,x\right)  $ is
the solution to the Kolmogorov's forward equation
\[
\frac{\partial H\left(  t,x\right)  }{\partial t}+\frac{\partial H\left(
t,x\right)  }{\partial x}\Delta=H\left(  t,x\right)  Q,
\]
with the boundary condition
\[
H\left(  t,0\right)  =\pi\left(  0\right)  \delta\left(  t\right)  ,
\]
and the initial condition%
\[
H\left(  0,x\right)  =\pi\left(  0\right)  \delta\left(  x\right)  ,
\]%
\[
\text{\ }\delta\left(  x\right)  =\left\{
\begin{array}
[c]{l}%
1,\text{ \ }x=0,\\
0,\text{ \ }x>0.
\end{array}
\right.
\]

\textbf{(ii) Computing the first passage time:} Let $\Gamma\left(
x\right)  $ be the first passage time of the cumulative power
consumption $\Phi\left(  t\right)$ arriving at a key power value $x$
as follows:
\[
\Gamma\left(  x\right)  =\min\left\{  t:\Phi\left(  t\right)  =x\right\}  .
\]
We write
\[
C\left(  t,x\right)  =P\left\{  \Gamma\left(  x\right)  \leq t\right\}  .
\]
It is clear that the event $\left\{  \Gamma\left(  x\right)  \leq t\right\}  $
is equivalent to the event $\left\{  \Phi\left(  t\right)  >x\right\}  $.
Hence we get
\[
C\left(  t,x\right)  =1-P\left\{  \Phi\left(  t\right)  \leq x\right\}
=1-\Theta\left(  t,x\right)  ,
\]
Hence, we have
\[
P\left\{  \Gamma\left(  x\right)  \leq t\right\}  =1-P\left\{  \Phi\left(
t\right)  \leq x\right\}  ,
\]

Let
\[
M\left(  r,k\right)  =\frac{\partial^{r}}{\partial s^{r}}\left[  \left(
Q-sI\right)  \Delta^{-1}\right]  ^{k}\mid_{s=0}\text{, \ \ }r,k\geq0.
\]
Then
\[%
\begin{array}
[c]{ll}%
M\left(  0,0\right)  =I, & \\
M\left(  r,0\right)  =0, & \text{\ }r\geq1,\\
M\left(  0,k\right)  =\left(  Q\Delta^{-1}\right)  ^{k}, & k\geq1,\\
M\left(  1,1\right)  =-\Delta^{-1}, & \\
M\left(  r,k\right)  =Q\Delta^{-1}M\left(  r,k-1\right)  -r\Delta^{-1}M\left(
r-1,k-1\right)  , & r,k\geq1.
\end{array}
\]
Therefore, we can provide expression for the $\emph{r}$th moment
$E\left[ \Gamma\left(  x\right)  ^{r}\right]  $ as follows:
\[
E\left[  \Gamma\left(  x\right)  ^{r}\right]  =\left(  -1\right)  ^{r+1}%
\pi\left(  0\right)  \sum_{k=0}^{\infty}\frac{x^{k}}{k!}M\left(
r,k\right) e, \ \ \ r\geq 1,
\]
where $\pi\left(  0\right)  $ is any initial probability vector of the QBD
process $Q$.

\subsection{Open problems}

From the above analysis, it is seen that discussing
more general group-server queues is interesting, challenging and
difficult. Thus it may be valuable to list several open problems for
the future study of group-server queues as follows:

\begin{itemize}
\item Setting up some suitable cost (or reward) functions, and provide and
prove existence and structure of bilateral threshold control by means of
Markov decision processes.

\item Establishing fluid and diffusion approximations for more general
group-server queues, and focus on how to deal with the residual service times
of those concentratively transferred customers to the buffer or to the left-side server
groups. In fact, the residual service times cause some substantial difficulties in model
analysis.

\item Constructing martingale problems or stochastic differential equations
for more general group-server queues.

\item Developing stochastic optimization and control, Markov decision
processes and stochastic game theory in the study of group-server queues.
\end{itemize}

\section{Simulation Experiments}

In this section, we design some simulation experiments for
performance evaluation of two different group-server queues: the
group-server loss queues, and the group-server queues with infinite
buffer. Specifically, we analyze the expected queue lengths, the
expected sojourn times and the expected virtual service times for
the two group-server queues.

In these following simulations, we use some common parameters for
the two different group-server queues: the group-server loss queues
and the group-server queues with infinite buffer, where there are
three server groups in each queueing system. To that end, we take
that $N=2$, $m_{0}=4$, $m_{1}=4$, $m_{2}=3$, $\mu_{0}=5$,
$\mu_{1}=4$, $\mu_{2}=3$, $K_{1}=K_{2}=3$ and $L_{1}=L_{2}=2$.

\textbf{(1) The group-server loss queues}

Fig.4 indicates how the expected customer number in the whole system depends on
the arrival rate $\lambda\in(15,45)$. It is seen from Fig.4 that the expected customer number in the whole system
strictly increase as $\lambda$ increases.
\begin{figure}[hptb]
\begin{center}
\includegraphics[width=3.6 in]{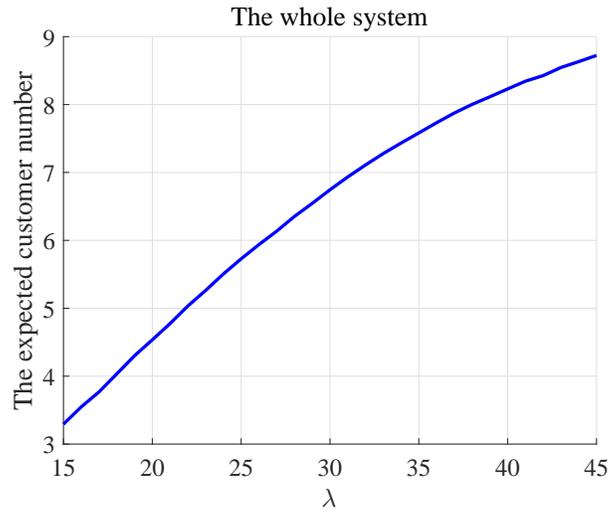}
\caption{The expected customer number in the whole system vs.
$\lambda$}
\end{center}
\end{figure}

Fig.5 shows how the expected customer number in each of three server groups depends on
the arrival rate $\lambda\in(15,45)$. It is seen from Fig.5 that the expected customer number in each of three server groups
strictly increases as $\lambda$ increases.

Note that each of the two group-server queues contains three
different server groups with service rates: $\mu_0, \mu_1, \mu_2$,
it is easy to see that the service times of the three server groups
are different from each other. In this case, for such a group-server
queue, we need to introduce a {\it virtual service time} as follows:
The virtual service time is defined as the average service time of any customer in the
group-server queues. Thus the virtual service time describes the
comprehensive service ability of a whole system through integrating
the server groups with different service abilities.
\begin{figure}[hptb]
\setlength{\abovecaptionskip}{0.cm}  \setlength{\belowcaptionskip}{-0.cm}
\centering                           \includegraphics[width=7.5cm]{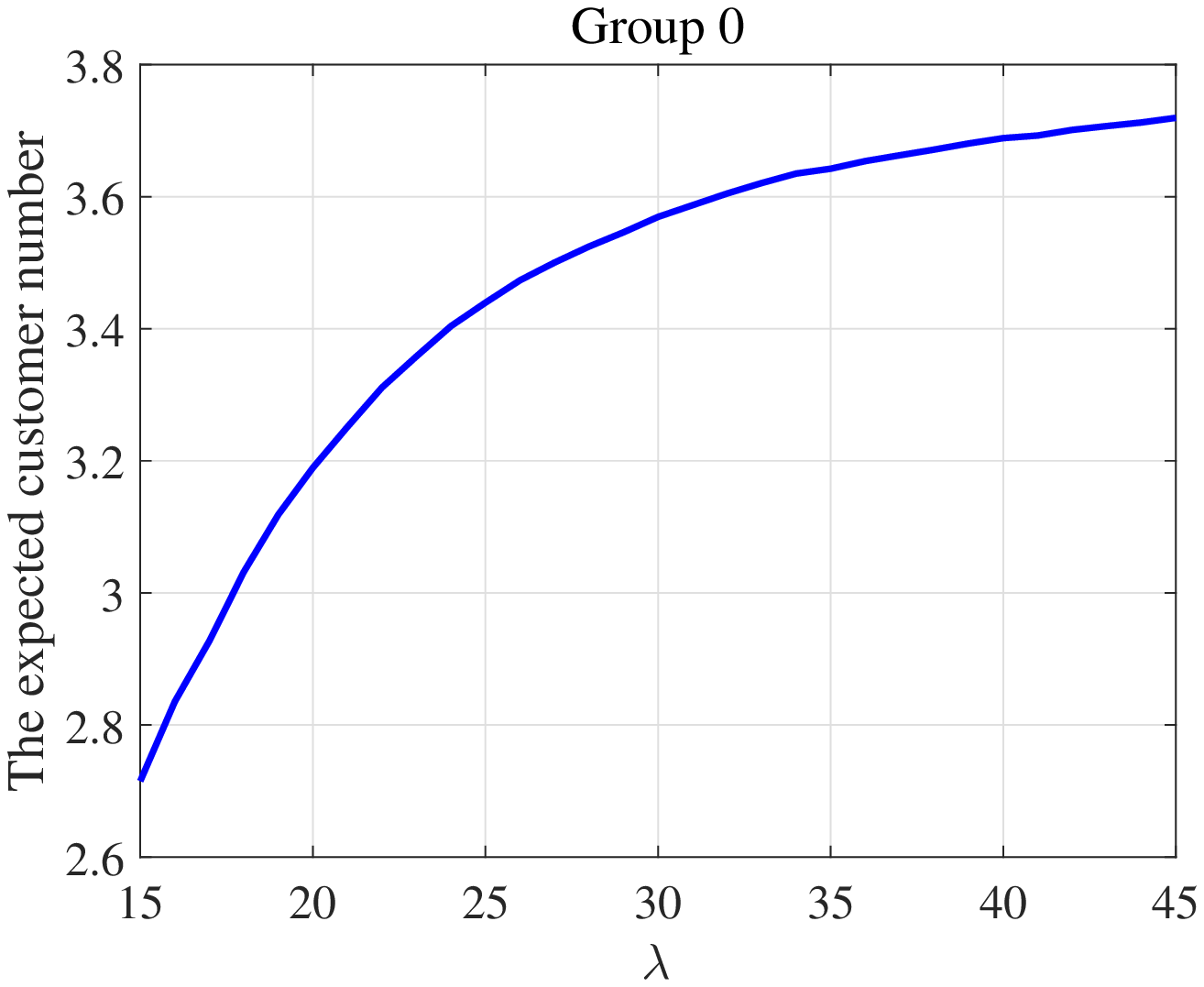}
\includegraphics[width=7.5cm]{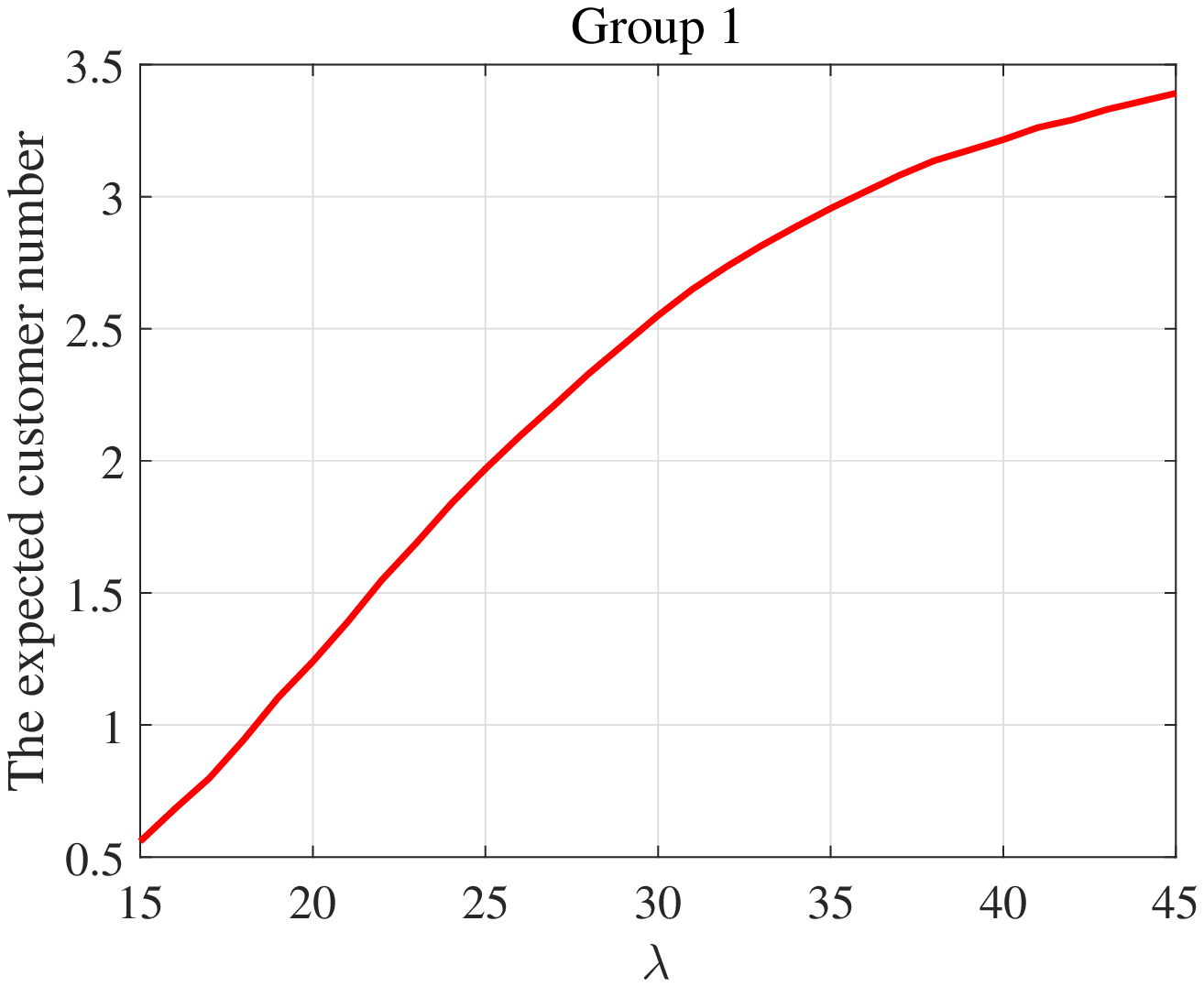}    \includegraphics[width=7.5cm]{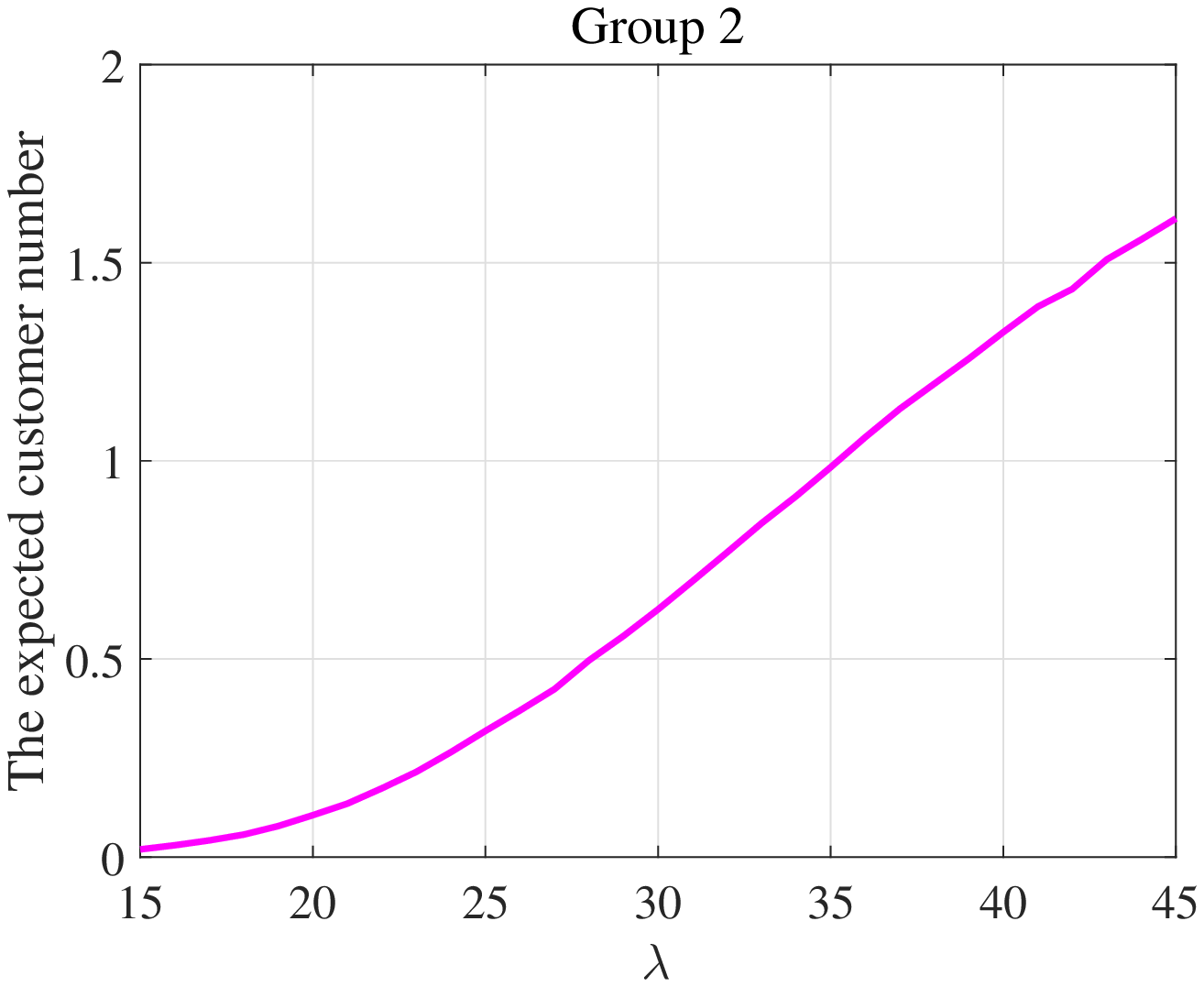}
\newline  \caption{The expected customer number in each of three server groups vs. $\lambda$}
\label{figure:fig-4}%
\end{figure}

Fig.6 demonstrates how the expected virtual service time and the expected sojourn time depend on
the arrival rate $\lambda\in(15,43)$, respectively. It is seen from Fig.6 that both of them
strictly increase as $\lambda$ increases.
\begin{figure}[hptb]
\begin{center}
\includegraphics[width=3.6in]{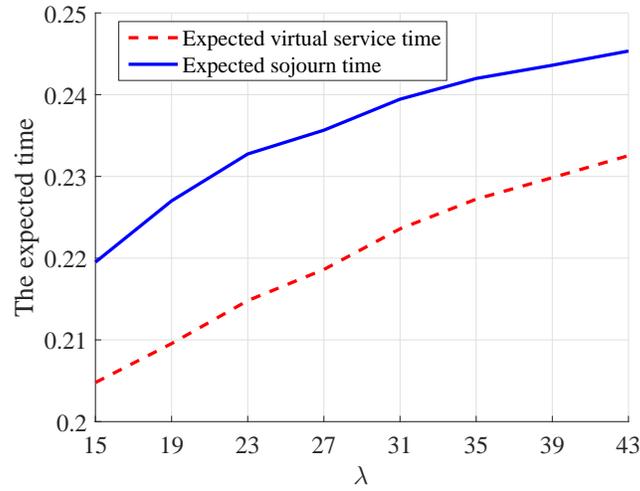}
\caption{The expected virtual service time and the expected sojourn time vs. $\lambda$}
\end{center}
\end{figure}

\textbf{(2) The group-server queues with infinite buffer}

Fig.7 indicates how the expected customer numbers in the whole system and in the buffer depend on
the arrival rate $\lambda\in(15,40)$, respectively. It is seen from Fig.7 that both of them
strictly increase as $\lambda$ increases.
\begin{figure}[htbp]
\begin{center}
\includegraphics[width=3.6 in]{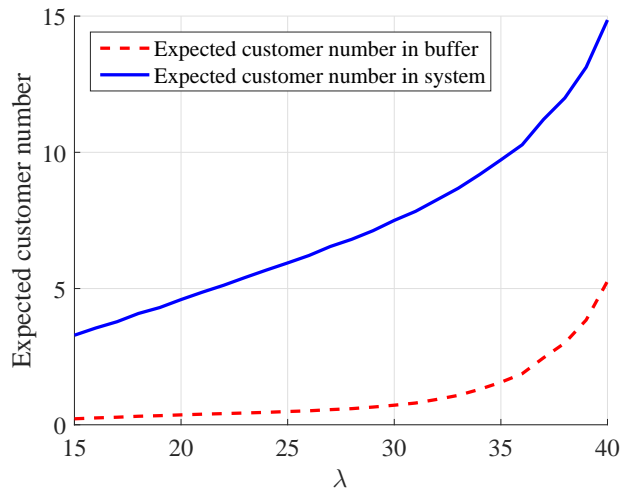}
\caption{The expected customer numbers in system and buffer vs.
$\lambda$}
\end{center}
\end{figure}

Fig.8 shows how the expected customer number in each of three server groups depends on
the arrival rate $\lambda\in(15,45)$. It is seen from Fig.8 that the expected customer
number in each of three server groups strictly increases as $\lambda$ increases.

\begin{figure}[hptb]
\setlength{\abovecaptionskip}{0.cm}  \setlength{\belowcaptionskip}{-0.cm}
\centering
\includegraphics[width=7.5cm]{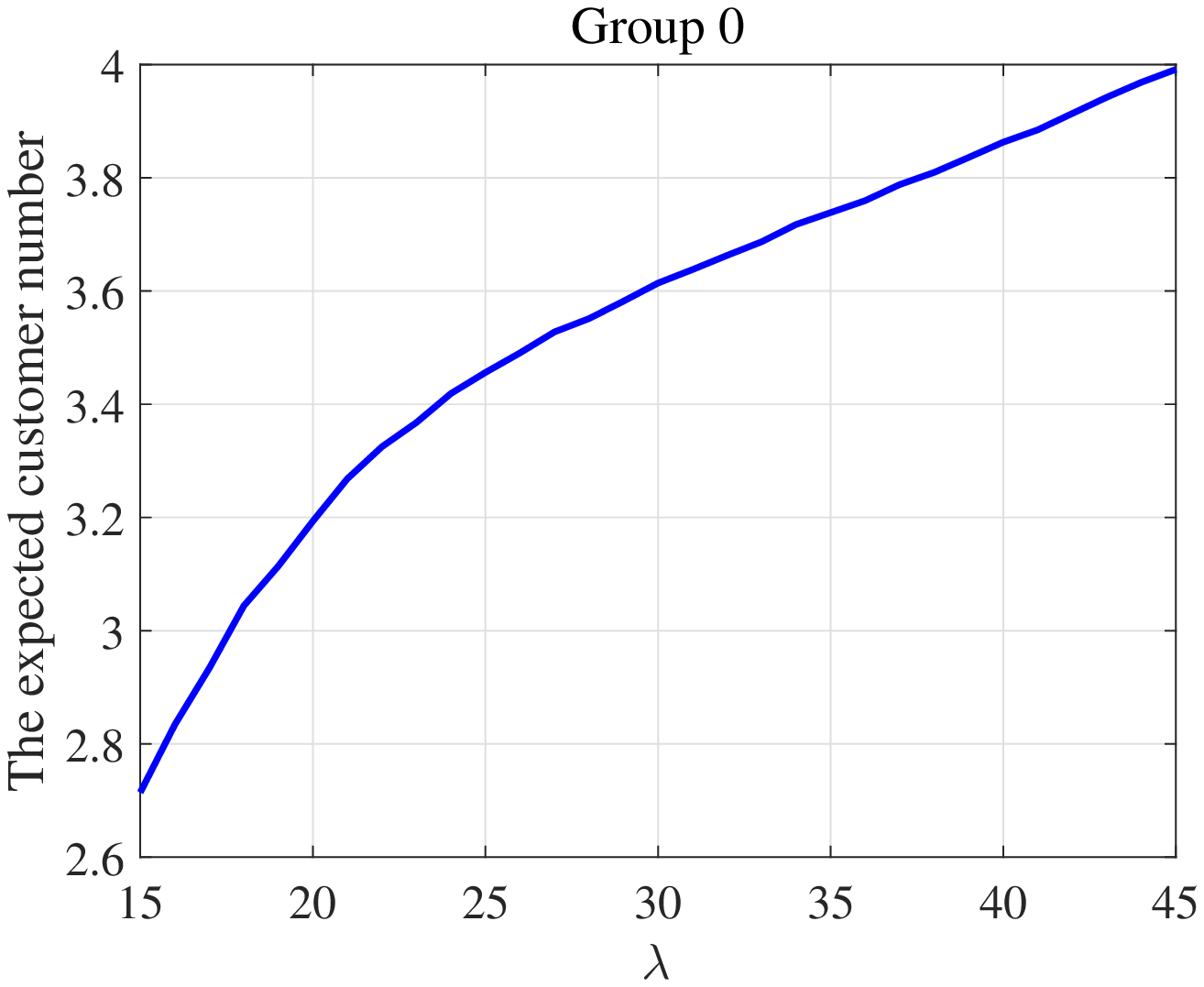}
\includegraphics[width=7.5cm]{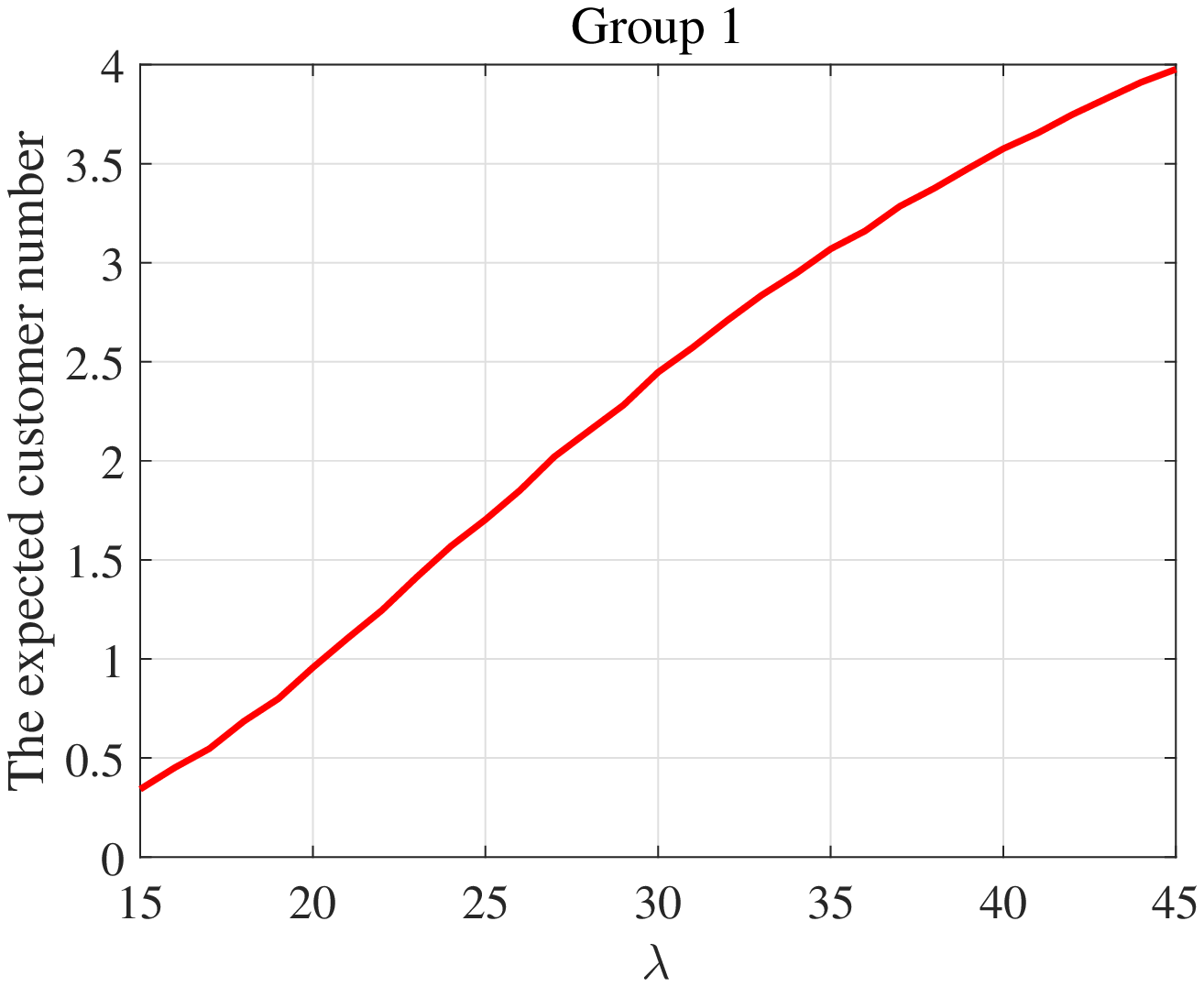}
\includegraphics[width=7.5cm]{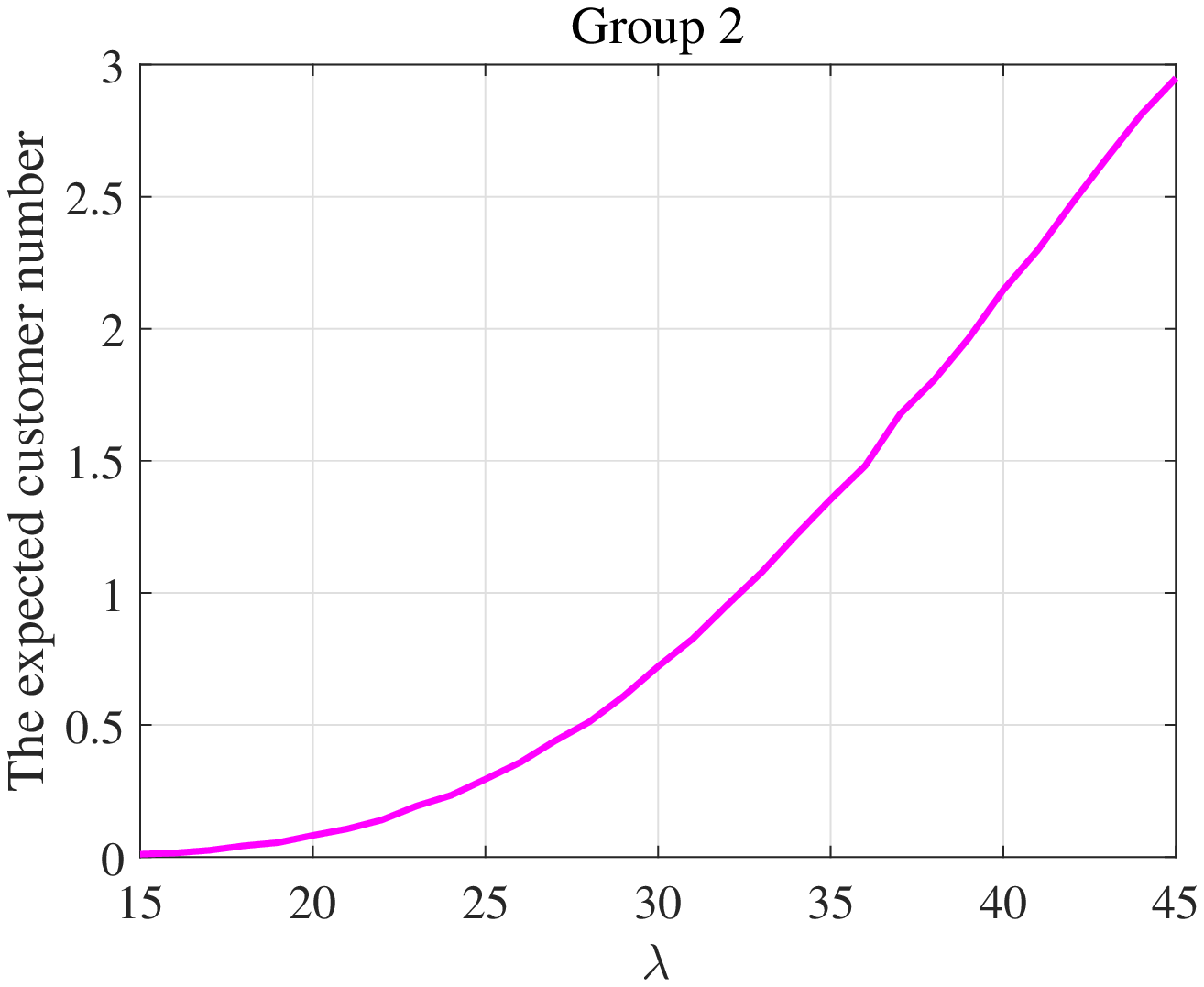}
\newline  \caption{The expected customer number in each of three server groups vs. $\lambda$}
\end{figure}
\begin{figure}[htbp]
\begin{center}
\includegraphics[width=3.6 in]{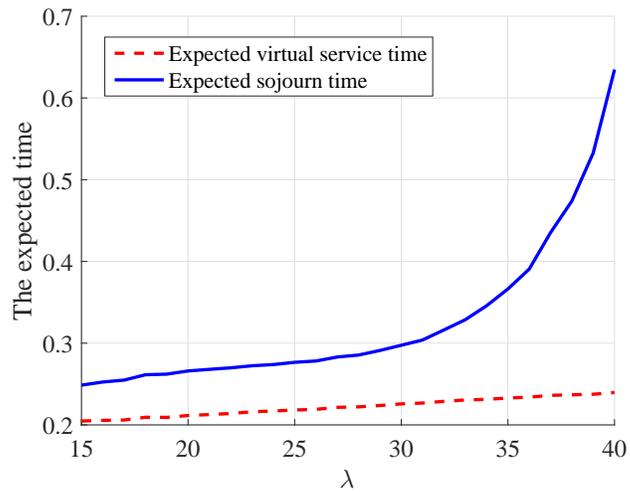}
\caption{The expected virtual service time and the expected sojourn time vs. $\lambda$}
\end{center}
\end{figure}

Fig.9 demonstrates how the expected virtual service time and the
expected sojourn time depend on the arrival rate
$\lambda\in(15,40)$, respectively. It is seen from Fig.9 that the
expected virtual service time increases slowly as $\lambda$
increases, while the expected sojourn time increases rapidly as
$\lambda$ increases.

Here, we use some simulation experiments to give valuable
observation and understanding with respect to system performance, and
this is a valuable help for design, operations and optimization of
energy-efficient management of data centers. Therefore, such a numerical
analysis will also be useful and necessary in the energy-efficient management
study of data center networks in practice.

\section{Concluding Remarks}

In this paper, we propose and develop a class of interesting {\it
Group-Server Queues} by means of analyzing energy-efficient
management of data centers, and establishes two representative
group-server queues through loss networks and impatient customers,
respectively. Some simple mathematical discussion is provided in the
study of two-group-server loss queues, and simulations are made to
study the expected queue lengths, the expected sojourn times and the
expected virtual service times in the three-group-server queues with infinite buffer. Furthermore, we show that this class
of group-server queues are often encountered in many other practical
areas including communication networks, manufacturing systems,
transportation networks, financial networks and healthcare systems.
Therefore, not only analysis of group-server queues is regarded as a
new interesting research direction, but there also exists many
theoretic challenge and basic difficulties in the area of queueing
networks. We hope the methodology and results given in this paper
can be applicable to analyzing more general large-scale data center
networks and service systems. Along these lines, we will continue
our works in following directions in the future research:

\begin{itemize}
\item Establishing fluid and diffusion approximations for the group-server loss queues, and also for
the group-server queues with impatient customers and with infinite
buffers;

\item making power consumption rate and power price regulation in data center networks through the Brownian
approximation methods;

\item setting up and proving existence and structure of (discrete or continue) bilateral threshold control in energy-efficient
management of data centers in terms of Markov decision processes; and

\item developing stochastic optimization and control, Markov decision processes and stochastic game theory in analyzing energy-efficient management of data center networks.
\end{itemize}

\section*{Acknowledgments}
This work was supported in part by the National Natural Science
Foundation of China under grant No. 71671158 and No. 71471160; and
by the Natural Science Foundation of Hebei province under grant No.
G2017203277.

\end{document}